\documentclass{jaa}
\usepackage{graphicx}
\usepackage{xcolor, soul}
\begin{document}\sloppy

%%paper title
%%For line breaks \\ can be used within title
\title{Low luminosity observation of BeXRB source IGR J21347+4737}

%%author names are separated by comma (,)
%%use \and before the last author name
%%use a * along with the number separated by comma
%% for the author for correspondence
%%\textsuperscript{number} is used for affiliation
%%\affilOne, \affilTwo etc., upto \affilTwentyfive is possible
%%Please note the first letter after \affil is capitalised in the command
%%

\author{Manoj Ghising\textsuperscript{1}, Ruchi Tamang\textsuperscript{1},  Binay Rai\textsuperscript{1}, Mohammed Tobrej\textsuperscript{1}, and Bikash Chandra Paul\textsuperscript{1}}
\affilOne{\textsuperscript{1}Department of Physics, North Bengal University,Siliguri, Darjeeling, WB, 734013, India\\}
\affilTwo{\textsuperscript{2}ICARD, Depatment of Physics, North Bengal University, Siliguri, Darjeeling, WB, 734013, India}

%%escape two column mode for title, affiliation and abstract
%%by giving \twocolumn command as shown

\twocolumn[{

\maketitle

%%include \corres to print the corresponding author Email id
\corres{bcpaul@associates.iucaa.in}

%%include \msinfo for
%%manuscript information such as
%%received, revised and accepted dates
%%

%%abstract
\begin{abstract}
In this paper, we report the results of the detailed temporal and spectral studies of the BeXRB J21347+4737 based on the data from the NuSTAR and \textit{SWIFT/XRT} in a wide energy range of 0.5-50 keV. Coherent pulsation with a period of 322.738$\;\pm\;0.018$ s was found in the light curve, implying, the source pulsation has spun down by 0.341 s $yr^{-1}$ when compared with the coherent pulsation estimated from XMM Newton more than 7 years ago.
 The pulse profile of the source demonstrates energy dependence and has evolved with time. The pulse fraction of the source observed by NuSTAR initially decreases with energy upto $\sim$15 keV, followed by a non-monotonic increasing trend above 15 keV. The source spectrum can be well approximated by an absorbed power-law model with modification by an exponential cutoff at high energies. The absorbed flux of the source is $4\times10^{-11}\;erg\;cm^{-2}\;s^{-1}$ and its corresponding luminosity is $3.5\times10^{35}\;erg\;s^{-1}$. The study of pulse-phase resolved spectroscopy shows a strong variation of spectral parameters on the phase. No additional emission or absorption features in the form of Fe line or Cyclotron lines were observed both in the phase-averaged and phase-resolved spectra of IGR J21347+4737.
\end{abstract}

%%insert keywords separated by 3 hyphens using \keywords{words}

}]
%%close the twocolumn escape here

%%include \doinum{number}for the DOI number in the header
%%include \volnum{number} for the volume number in the header
%%include \year{yyyy} for year of publication in the header
%%include \pgrange{num--num} page range of article in the header
%%include \artcitid{num} for the article citation id
%%include \lp to print last page of the article
%%include \setcounter{page}{pagenum} for the exact starting page of the article

\doinum{12.3456/s78910-011-012-3}
\artcitid{\#\#\#\#}
\volnum{000}
\year{0000}
\pgrange{1--}
\setcounter{page}{1}
\lp{1}

\section{Introduction}
High Mass X-ray Binaries (HMXBs) are binary systems consisting of an early-type massive star denoted as the donor star and a compact object that is either a black hole or a neutron star. Multi-wavelength studies of HMXBs provide exceptional astrophysical laboratories in understanding stellar evolution, accretion physics, and gravitational wave events. HMXBs are categorized into two classes \textit{viz.} Be/X-ray Binaries (BeXRBs) and Super Giant X-ray Binaries (SGXBs) (Reig et al. 2011). The BeXRB system is known for harboring a neutron star and a fast spinning early-type star with an equatorial circumstellar disc (Reig
et al. 2011). Be stars are a subset of B-type stars, non-supergiant fast rotating, luminosity class III-IV stars, which at some point in past have shown spectral lines in emission (Porter \&
Rivinius 2003). These systems represent an extreme case of X-ray variability spanning up to four orders of magnitude.  They are most of the time in a quiescent state and show transient character. They are best studied by various observatories during bright outbursts when the count rate is significantly enhanced. These systems undergo two types of outbursts- Type I and Type II (Reig \& Nespoli 2013). However, the luminosity in the case of Type-II outburst activity is 1-2 orders of magnitude more than Type-I outbursts. Type-II outbursts are less frequent compared to Type-I which are periodic, function of the orbital period. The type-II outbursts last relatively longer than Type-I and last for a large fraction of an orbital period or several orbital periods (few sources reported undergoing Type-II outbursts are EXO 2030+375 (Wilson, Finger \&
Camero 2008) and GRO J2058+42 (Molkov et al. 2019) while Type-I covers a small fraction of the orbital period (0.2-0.3 $P_{\odot}$) (few sources reported undergoing Type-I outbursts are LXP 38.55 (Vasilopoulos et al. 2016a), GX 304-1 (Jaisawal et al. 2016).  

The BeXRB IGR J21347+4737 was discovered using INTEGRAL/IBIS in 2002 (Krivonos 2007; Bird 2007). Chandra observations (Sazonov et al. 2008) helped in determining accurate source position and which allowed identification of optical counterpart using observations in the optical band; the optical counterpart has also been observed by Masetti et al. (2009) with an optical spectral appearance completely different from the findings of Bikmaev et al. (2008). Masetti et al. (2009) report on a source distance of about 5.8 kpc. The optical counterpart is present in the Gaia DR3 catalogue (Gaia DR3 1978365123143522176), where it has a distance of 5.1 kpc. An important property of this star is that it is a shell Be star, implying that it is observed almost edge on (e.g. Reig \& Fabregat 2015). The coherent pulsation of the source was first reported at 320.35 s by Reig \& Zezas (2014). During the all-sky X-ray monitoring campaign of IBIS, the source underwent from the active state (2002 Dec - 2004 Feb) to inactivity (2004 Mar. - 2007 Feb.). The average X-ray flux (17-60 keV) of the source during the period of activity was (2.3 $\pm\; 0.4)\times\;10^{-11}\;erg\;cm^{-2}\;s^{-1}$ (Bikmaev et al. 2008). The  source flux was reported to have increased to 7$\;\times\;10^{-11}\;erg\;cm^{-2}\;s^{-1}$ in the 4-12 keV band in comparison to 1.3 $\times\;10^{-11}\;erg\;cm^{-2}\;s^{-1}$ during the second all-sky survey of ART-XC telescope onboard the SRG observatory. The SRG/ART-XC reported a possible beginning of a new outburst from the source, however, later observations of the NuSTAR about 14 days after found that the source flux decreased to 1.49 $\times\;10^{-11}\;erg\;cm^{-2}\;s^{-1}$, which hinted that the source was not entering the outburst state.
 
In this paper, we probe the detailed coverage of X-ray timing and spectral properties of the BeXRB IGR J21347+4737 in the broadband 0.5-50 keV energy range. The \textit{Swift} and NuSTAR observation has been considered for analysis.

\subsection{NuSTAR}

The Nuclear Spectroscopic Telescope Array (NuSTAR), is a NASA space-based X-ray telescope for studying high-energy X-rays from astrophysical sources. It was the first hard X-ray focussing telescope that operates in the energy range (3-79) keV and consists of two identical X-ray telescope modules that have their own Focal plane modules referred as \textsc{fpma} and \textsc{fpmb} (Harrison et al. 2013). The telescope provides an X-ray imaging, timing, and spectroscopy with the angular resolution of 18 arcsec and spectral resolution of 400 eV at 10 keV. The light curves and the spectra were analyzed using the latest version of \textsc{heasoft } \textit{v6.30.1}. The data reduction of the source IGR J21347+4737 was processed using the standard software \textsc{nustardas} v1.9.7. In order to get the clean event files for the respective modules, we run the mission-specific task \textsc{nupipeline}. Using the standard \textit{ftool} \textsc{xselect} combined with the \texttt{SaoImage} \textsc{ds9} application software, the source and background regions were selected where a circular region of 80" around the source center was considered as the region file for the source and the background region of the same size was taken away from the source as the background region file. The light curve file and the spectra file were obtained by running the script \textsc{nuproducts} making use of the region files obtained earlier. The background subtraction of the light curve for both the instruments \textsc{fpma} and \textsc{fpmb} were done by using \textit{ftool} \textsc{lcmath} and finally, barycentric corrections were performed with the help of \textit{ftool} \textsc{barycorr}.      

\subsection{XMM-Newton}
The XMM-Newton data were extracted so that we could conduct a comparative analysis in the soft energy band. The XMM–Newton \textsc{epic} instrument data are reduced by using the Science Analysis System software \textsc{sas} \textit{version 20.0.0}. In order to minimize the pile-up effect, both the \textsc{mos} and \textsc{pn} data were taken in small window mode. The \textsc{epic} date were screened and filtered using the script \textsc{epchain} for \textsc{pn}-detector while \textsc{emchain} for \textsc{mos 1} and \textsc{mos 2} detectors. We have excluded all events at the edge of \textsc{ccd} and from bad pixels by setting \textsc{flag=0} and selecting the pn events with \textsc{pattern} in the range 0-4, and the \textsc{mos} data with \textsc{pattern} $\leq$12. The source photons are extracted by considering a circular aperture of radius 25 arcsec, and the background is also taken from the same \textsc{ccd} chip with a circular region of radius 25 arcsec. Light curves at different energies were extracted by using the concatenated and calibrated \textsc{epic} event available in the \textsc{pps} pipeline products. The light curves were barycentred to the solar frame by using the task \textsc{barycen}. 

\subsection{\textit{Swift}}
The \textit{Swift/X-ray Telescope(XRT)} data was utilised to understand the spectral fit in the soft energy state (1-10) keV. The \textit{Swift} observatory also includes other two instruments \textit{viz.} the Burst Alert Telescope (BAT) and UV/Optical Telescope (UVOT) but not considered in the present manuscript. The combined three instruments XRT, BAT and UVOT covers an energy range of ~0.002-150 keV. The observatory \textit{Swift} performs regular monitoring in the energy range ~(15-50) keV of the X-ray sky (Krimm et al. 2013). The \textit{Swift/XRT} operates in the energy range 0.5-10 keV. Its data were extracted by imposing the mission-specific task \textsc{xrtpipeline}. The event files was extracted by using \textit{ftool} \textsc{xselect}. The extracted image was viewed by using the astronomical imaging software Ds9. A circular region of 30 arcsec and background region of the same size was considered for observations in photon counting mode. 
 
\begin{table*}
 \begin{center}
 \begin{tabular}{clllc}
    \hline
    \hline
    
   Observatory	& Date of observation &	OBs ID	&	Exposure & Count Rates	$(cs^{-1})$\\
	&		&	& (in ksec)	 & \\
\hline	
\hline
			
NuSTAR	   &	2020-12-17 & 90601339002	&	27.10 &	1.54$\pm0.01$ \\
\textit{Swift/XRT} & 2020-12-17           &  00089189001            &       1.63 &  0.077$\pm0.006$       \\
XMM-Newton &    2013-11-24  & 0727961301     &   30    & 0.14$\pm0.02$ \textsc{mos1}\\
		   &	           &	            &	      &	0.13$\pm0.01$ \textsc{mos2} \\
		   &	           &	            &	      &	0.38$\pm0.03$ \textsc{pn}\\
    
      \hline
      \hline
  \end{tabular}
  \caption{Observation details of the source IGR J21347+4737.}
  \end{center}
 \end{table*}

\section{TIMING ANALYSIS}

\begin{figure}

\begin{center}
\includegraphics[angle=0,scale=0.3]{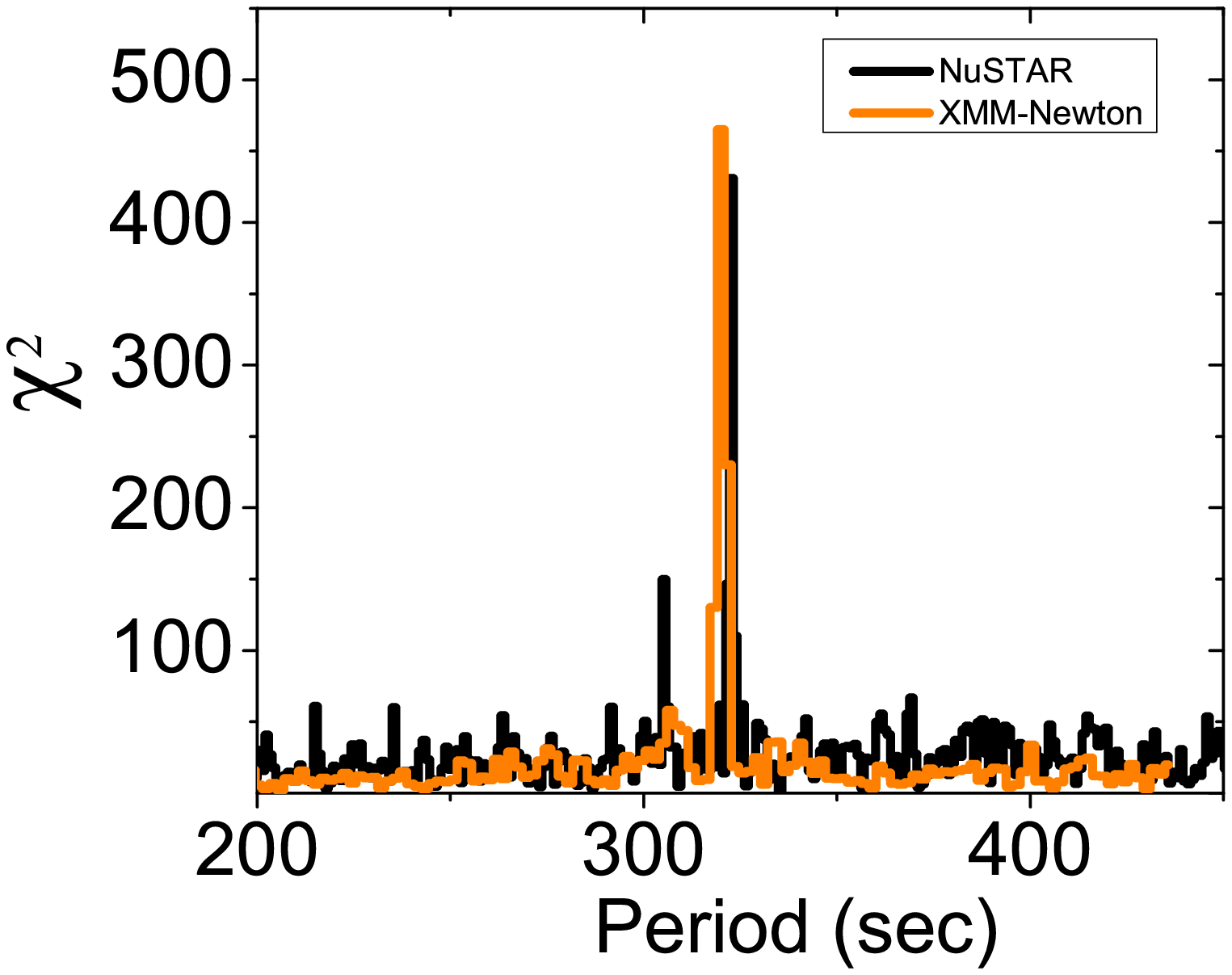}
\end{center}
\caption{Periodogram of the source IGR J21347+4737 corresponding to both NuSTAR (black) and
XMM-Newton (orange) observation. }

\end{figure}

\subsection{Pulse period estimation}
The source and background light curves with a bin size of 2 s were considered for \textit{NuSTAR} for pulse period estimation. An approximate pulse period of the source was detected by using \textsc{heasarc} task \textsc{powspec} and the precise value were determined by making use of epoch folding technique (Leahy et al. 1983) through the \textsc{efsearch} tool. The precise pulse period of the source IGR J21347+4737 were detected at 322.738$\pm0.018$ s. The uncertainties in the spin period were estimated by simulating light curves following the methods outlined in (Boldin et al. 2013). For this, we generated 1000 simulated light curves such that the count rates are within the error of the original data. Next, we applied the epoch folding technique to each simulated light curve to obtain the spin period distribution. The mean value and the standard deviation of the distribution were then obtained. The standard deviation thus obtained were taken as the spin period uncertainty. 

In order to develop an understanding of the spin evolution of the source, we carried out the same analysis as above for \textit{XMM-Newton} observations were taken for more than 6 years back ago. The analysis of the light curve (bintime 40 s) estimates the spin period of the source at 320.351$\pm0.022$ s. Therefore, it turns out that in more than 7 years gap interval the pulsations of the source have spun down by 1.08$\times10^{-8} ss^{-1} $ consistent with the works of Pike et
al. (2020). A combined plot of the periodogram has been furnished in the Figure 1 for more clarity. 

\subsection{Pulse Profile and Pulse Fraction} 

\begin{figure*}

\begin{center}
\includegraphics[angle=0,scale=0.6]{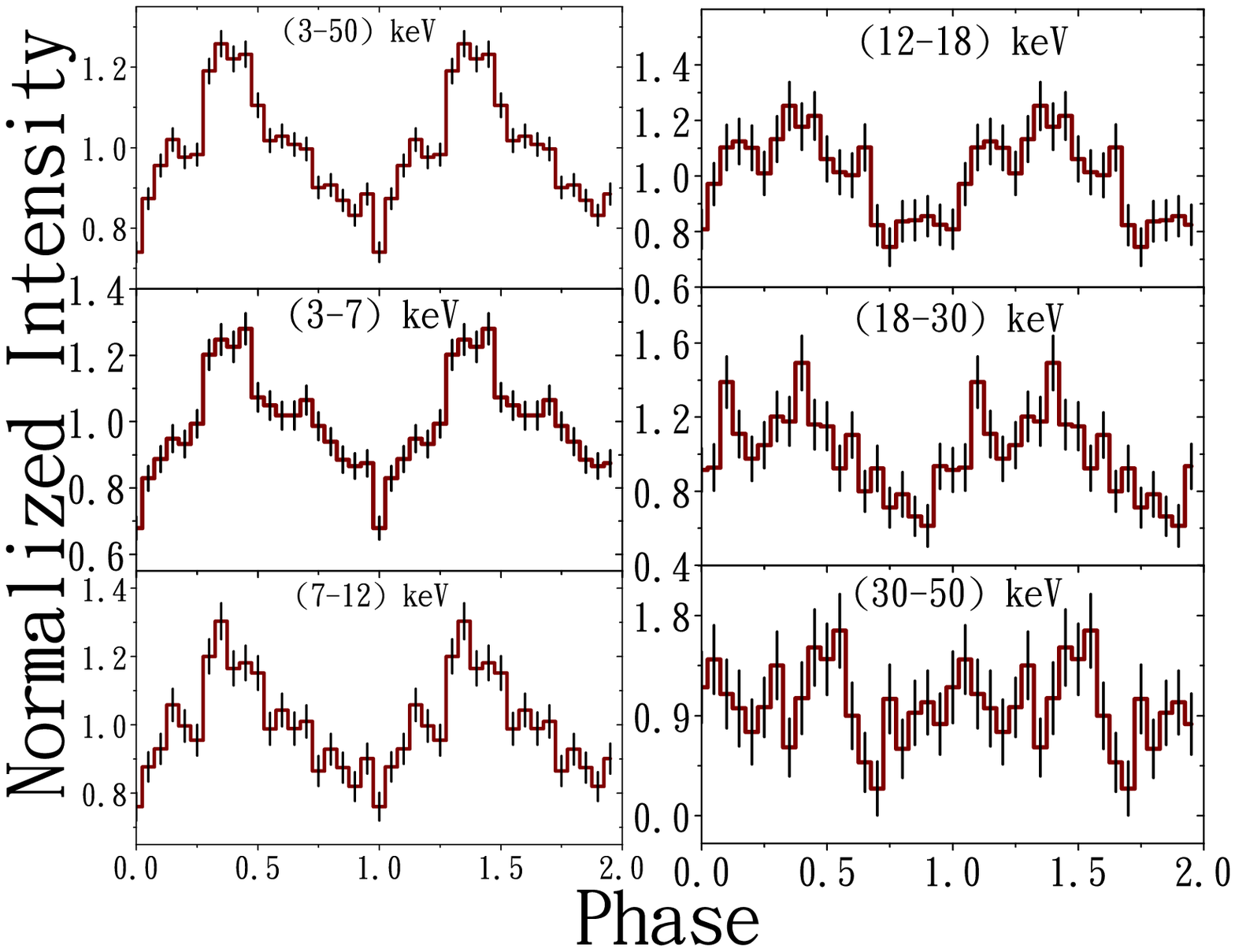}
\end{center}
\caption{Folded pulse profile of IGR J21347+4737 in several energy bands corresponding to NuSTAR observation.}
\end{figure*}

\begin{figure}

\begin{center}
\includegraphics[angle=0,scale=0.3]{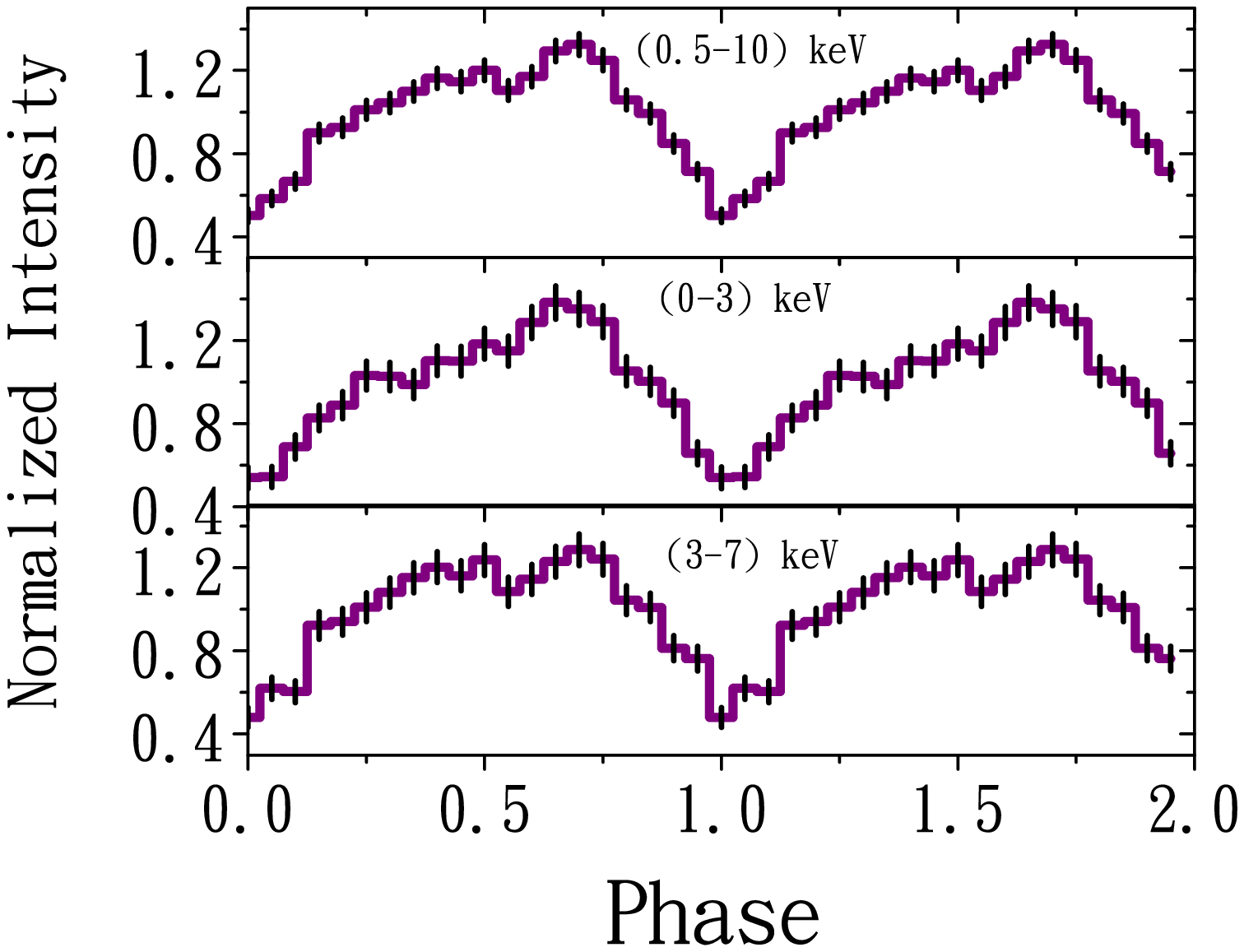}
\end{center}
\caption{Folded pulse profile of IGR J21347+4737 in several energy bands corresponding to XMM-Newton observation.}
\end{figure}

\begin{figure}

\begin{center}
\includegraphics[angle=0,scale=0.3]{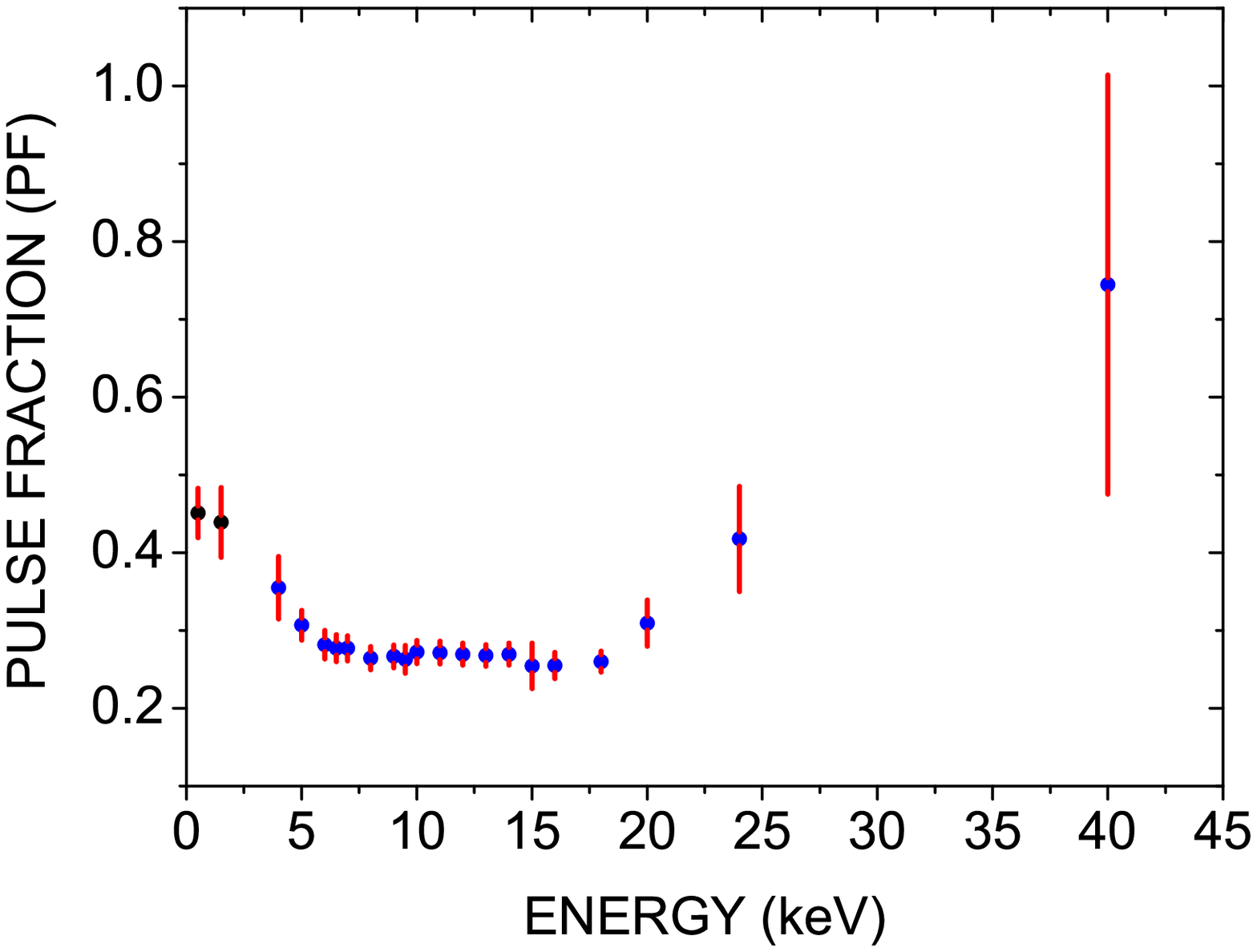}
\end{center}
\caption{Variation of Pulse Fraction (PF) with energy of IGR J21347+4737. The two symbols marked in black color correspond to XMM-Newton observations while blue color corresponds to NuSTAR observation.}
\end{figure}

\begin{figure*}

\begin{minipage}{0.3\textwidth}
\includegraphics[height=1.3\columnwidth, angle=0]{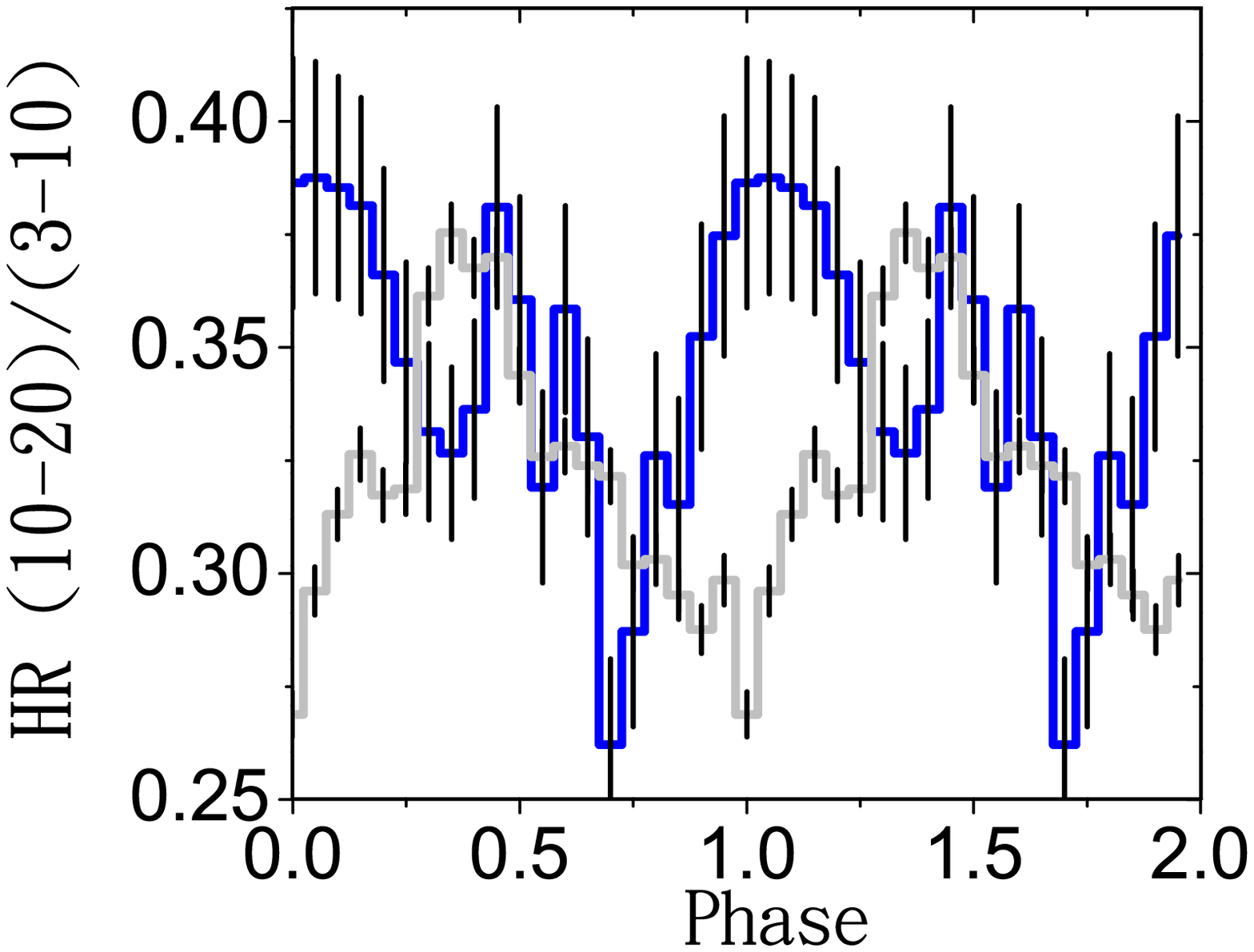}
\end{minipage}
\hspace{0.20\linewidth}
\begin{minipage}{0.3\textwidth}
\includegraphics[height=1.3\columnwidth, angle=0]{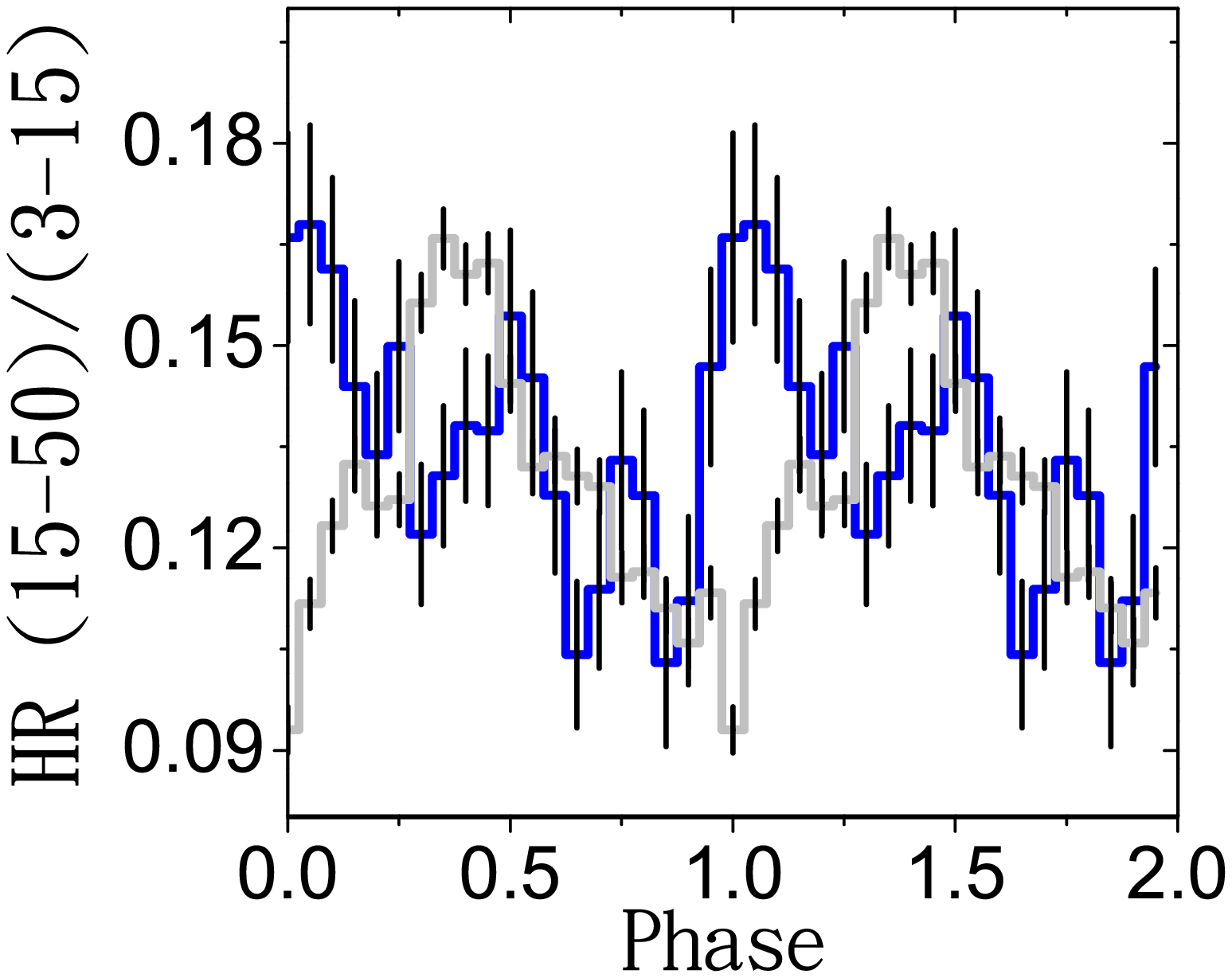}
\end{minipage}
\caption{Variation of Hardness Ratio (HR) for the IGR J21347+4737 pulse profiles as a function of the spin phase. The averaged pulse profile in a wide energy range 3-50 keV is superimposed in gray color for visual comparison.  }
\label{fig-2}
\end{figure*}

The pulse profile of the source was analyzed by resolving the NuSTAR light curves in 3-79 keV energy range into several energy bands of 3-7 keV, 7-12 keV, 12-18 keV, 18-30 keV, and 30-50 keV respectively. A similar analysis of XMM-Newton light curves was carried out in 0-3 keV, 3-7 keV, and 0.5-10 keV respectively. The pulse profiles shown in Figures 2 and 3 have been folded by defining the zero-point at the minimum flux. The pulse profile of both the observations is in general single-peaked and asymmetric demonstrating weak dependence on energy. Owing to limited statistics, no significant variations in higher energy bands is observed. The peak emission in the lowest energy band lies in the phase interval 0.4-0.5 with an additional emission component at phase interval 0.6-0.7. When compared to XMM-Newton observations, for e.g, the 3-7 keV energy range pulse profile, the peak emission has shifted towards lower phase intervals with time. This firmly establishes that the pulse profile has evolved with time.  

The Pulse Fraction (PF) is defined as the ratio of the difference between the maximum and minimum intensity ($F_{max}-F_{min}$) to the sum of the maximum and minimum intensity ($F_{max}+F_{min}$) of the pulse profile i.e $PF\;=(F_{max}-F_{min})/(F_{max}+F_{min})$. If we ignore XMM-Newton data, then pulse fraction of the NuSTAR observation initially decreases upto 15 keV and then above 15 keV, it increases with energy which is quite typical of these X-ray pulsars (Lutovinov \& Tsygankov 2009).

\subsection{Hardness Ratio (HR)}
The HR is defined as the ratio of unnormalized pulse profiles in the corresponding energy bands 10-20 keV/ 3-10 keV and 3-15 keV/ 15-50 keV respectively. It is observed from the Figure 5 that the HR exhibits two peak structures with broad maxima at the initial phase $\sim$0.1 and a narrow one at $\sim$0.5. The HR shows a significant anti-correlation with the continuum pulse profile.   

\section{Spectral Analysis}
\begin{figure}

\begin{center}
\includegraphics[angle=270,scale=0.3]{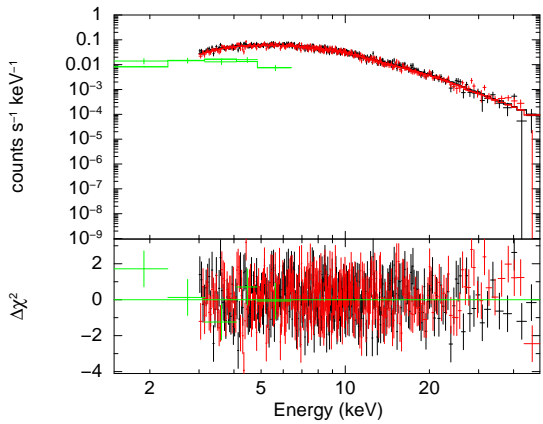}
\end{center}
\caption{Folded spectrum of IGR J21347+4737 and its approximation with the model \textsc{constant$\times$tbabs$\times$cutoffpl} in the energy range 0.5-50keV. Red
and black colors are for the FPMA and FPMB telescopes of the NuSTAR observatory while green color corresponds to \textit{Swift/XRT}.}
\end{figure}

\begin{figure}

\begin{center}
\includegraphics[angle=270,scale=0.3]{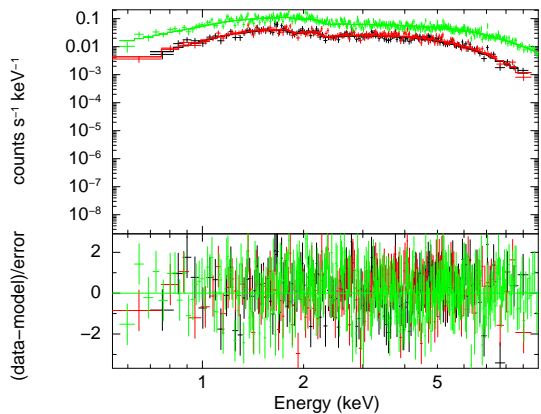}
\end{center}
\caption{Folded spectrum of XMM-Newton in the energy range 0.5-10 keV. The above Figure shows the spectrum of IGR J21347+4737 and its approximation with the model \textsc{constant$\times$tbabs$\times$(powerlaw+blackbody)}. Red
and black colors corresponds to MOS 1 and MOS 2 while green color is for pn.}
\end{figure}

\begin{table*}
\begin{center}
\begin{tabular}{clllllllllllc}
\hline	
\hline
Parameters	&	&		MODEL I (CUTOFFPL)	&	&		MODEL II (HIGHECUT)		\\
\hline													
$C_{FPMA}$	&	&		1(fixed)	&	&		1(fixed)		\\
$C_{FPMB}$	&	&		1.003$\pm$0.013	&	&		1.003$\pm$0.013		\\
$C_{XRT}$	&	&		0.84$\pm$0.02	&	&		0.84$\pm$0.03		\\
$n_{H}\;(\times\;10^{22}\;cm^{-2})$	&	&		 4.37$\pm$0.54	 & & 5.03$\pm$0.92	\\
$\Gamma$	&	&		1.30$\pm$0.07	&	&		1.34$\pm$0.09			\\
$E_{CUT}$ (keV)	&	&		20.06$\pm$2.26	&	&		5.98$\pm$0.69			\\
$E_{fold}$ (keV)	&	&		-	&	&		21.78$\pm$3.08			\\
Flux($\times   10^{-11}\;erg\;cm^{-2}\;s^{-1}$)	&	&		4.13$\pm$0.09	&	&		4.17$\pm$0.08		\\
$\chi_{\nu}^{2}$	&	&		1.09	&	&		1.08		\\

\hline
\hline
  \end{tabular}
  \caption{The above table represents the fit parameters of IGR J21347+4737 of \textit{Swift/XRT} and NuSTAR observations in the energy range 0.5-50 keV using the continuum models \textsc{constant$\times$tbabs$\times$cutoffpl} and \textsc{constant$\times$tbabs$\times$highecut$\times$po}. $n_{H}$ represents equivalent Hydrogen column density, $\Gamma$ and $E_{CUT}$ represents Photon Index  and \textsc{cutoff} energy of \textsc{cutoffpl} model and $E_{fold}$ represents folded energy of \textsc{highecut} model. Flux were calculated within energy range (0.5-50) keV. The fit statistics $\chi_{\nu}^{2}$  represents reduced $\chi^{2}$ ($\chi^{2}$ per degrees of freedom). Errors quoted for each parameter are within 90\% confidence interval.}
  \end{center}
 \end{table*}

\begin{table}
\begin{center}
\begin{tabular}{clllllllllllc}
\hline
\hline	
Parameters	&	&	&	XMM-Newton Data		\\
\hline
\hline										
$C_{MOS1}$	&	&	&	1(fixed)		\\
$C_{MOS2}$	&	&	&	1.06$\pm$0.0.04		\\
$C_{pn}$	&	&	&	1.03$\pm$0.0.03		\\
$n_{H}\;(\times\;10^{22}\;cm^{-2})$	&	&	&	 0.96$\pm$0.07		\\
bbodyrad (kT) (keV)	&	&	&	1.70$\pm$0.19		\\
bbodyrad norm	&	&	&	0.019$\pm$0.004		\\
$\Gamma$	&	&	&	0.81$\pm$0.12		\\
Flux ($\times  \; 10^{-12}\;erg\;cm^{-2}\;s^{-1}$)	&	&	&	3.85$\pm$0.16		\\
$\chi_{\nu}^{2}$ 	&	&	&	1.10		\\
\hline										
\hline
  \end{tabular}
  \caption{The above table represents the fit parameters of IGR J21347+4737 for XMM-Newton observations using the continuum model \textsc{constant$\times$tbabs$\times$(powerlaw+blackbody)}. $n_{H}$ represents equivalent Hydrogen column density, $\Gamma$ represents Photon Index and kT represents blackbody temperature. Flux were calculated within energy range (0.5-10) keV. The fit statistics $\chi_{\nu}^{2}$  represents reduced $\chi^{2}$ ($\chi^{2}$ per degrees of freedom). Errors quoted for each parameter are within 90\% confidence interval.}
  \end{center}
 \end{table}

\subsection{Phase-average spectroscopy}

The combined \textit{Swift}-NuSTAR phase-average spectra of the BeXRB IGR J21347+4737 were fitted in the energy range (0.5-50) keV (see Figure 6). We have ignored the spectrum below 3 keV for NuSTAR because the instrument is well calibrated only above 3 keV, and above 50 keV due to background domination. The \textit{Swift/XRT} spectra has been considered in the energy range 0.5-10 keV in order to understand the spectrum missed by NuSTAR below 3 keV. The spectra of both the FPMA and FPMB were grouped to have at least 30 counts per bin by using the tool \textsc{grppha}. The normalization factor \textit{i.e} inaccuracies in calibration between \textit{Swift} and NuSTAR were ensured by introducing the cross-calibration multiplicative factors (the \textsc{constant} model in \textsc{xspec}). The constant parameter of FPMA was fixed to unity while the parameter for FPMB and \textit{Swift} was left free. Several phenomenological models were applied that are broadly used to approximate the spectra of X-ray pulsars. In particular, the broadband spectra can be best fitted by a simple \textsc{cutoffpl} model (\textsc{model i}). The contribution of neutral hydrogen column density $(n_{H})$ was modified by the photoabsorption model (\textsc{tbabs} in \textsc{xspec}) with the solar abundances from (Wilms et al. 2000). No additional modifications with soft blackbody  were required to improve the spectra. Different continuum components in the form of \textsc{powerlaw}$\times$\textsc{highecut} \textsc{(model ii)} in place of \textsc{cutoffpl} were tested and they all fit the spectra very well with similar quality, however, the \textsc{cutoffpl} continuum makes it slightly better (see Table 2). The spectrum of the source has a typical shape for X-ray pulsars (Coburn et al. 2002;
Filippova et al. 2005). None of the model combinations revealed a significant presence of the Fluorescent iron line (Fe $K_{\alpha}$ line). The absorbed flux of the source in the 0.5-50 keV energy range was found to be $\sim4\times10^{-11}\;erg\;cm^{-2}\;s^{-1}$ and the corresponding luminosity is $3.45\times10^{35}\;erg\;s^{-1}$ assuming a distance of 8.5 kpc (Reig \& Zezas 2014). 

For comparative study and to understand the evolution of spectral parameters with luminosity and time, we have fitted the XMM-Newton observations in 0.5-10 keV. We have considered the joint fitting of \textsc{mos1}, \textsc{mos2}, and \textsc{pn} in our analysis (see Figure 7). The calibration uncertainty was ensured by keeping the constant parameter of \textsc{mos1} fixed while keeping free for the other two instruments. We have suitably fitted the spectra by an absorbed power-law model with an addition of bbodyrad component \textit{i.e.} \textsc{constant$\times$tbabs$\times$(powerlaw+blackbody)}. The details of the spectral parameters can be seen in Table 3. The absorbed flux of the source in the 0.5-10 keV energy range was found to be $\sim4\times10^{-12}\;erg\;cm^{-2}\;s^{-1}$ and the corresponding luminosity is $3.45\times10^{34}\;erg\;s^{-1}$ assuming a distance of 8.5 kpc (Reig \& Zezas
2014). 
\subsection{Phase-Resolved Spectroscopy}

\begin{figure}

\begin{center}
\includegraphics[angle=0,scale=0.3]{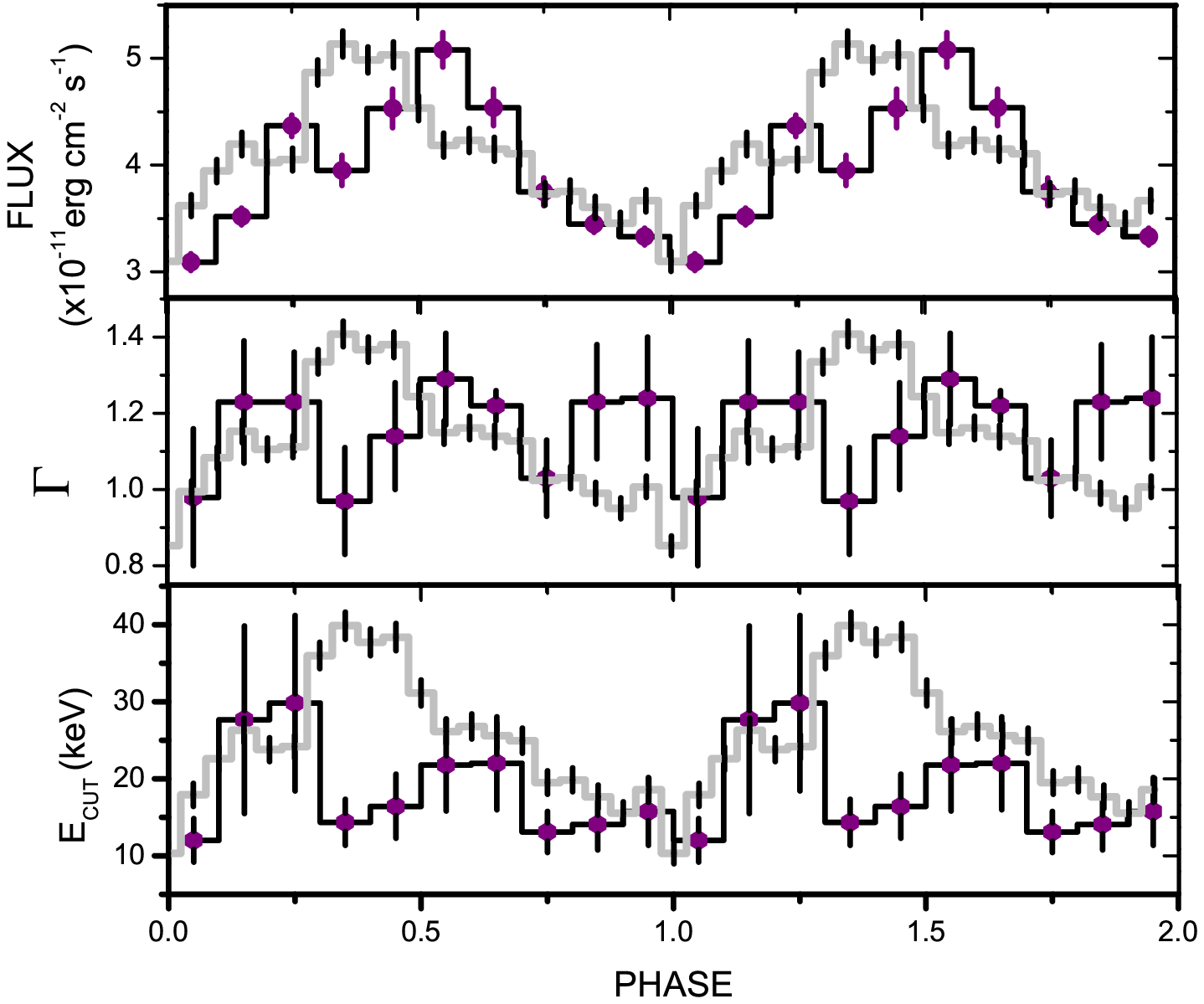}
\end{center}
\caption{Variation of spectral parameters with pulse phase. $\Gamma$ and $E_{CUT}$ represents Photon Index and \textsc{cutoff} energy of \textsc{cutoffpl} model. Flux were calculated within energy range (3-50) keV for the NuSTAR observation.
Errors quoted for each parameter are within 90 \% confidence interval. The averaged pulse profile in a wide energy range 3-50 keV is superimposed in gray color for visual comparison.}
\end{figure}

 In order to understand the evolution of the spectral parameters as a function of the rotation phase, we performed phase-resolved spectral analysis of the source in this section. Therefore, the source pulse period was divided into 10 evenly distributed rotational phase bins. We created \textsc{good time interval (gti)} files based on the folding epoch and rotational period corresponding to each phase. Using gti files merged by \textsc{mgtime}, we ran \textsc{nuproducts} for each phase to get the final spectra files corresponding to 10 phases.

The 3-50 keV spectra  corresponding to all the 10 phases were approximated with the best-fit model \textsc{constant$\times$tbabs$\times$cutoffpl} used for the phase-averaged spectrum. Figure 8 shows an evolution of the spectral parameters Flux, $\Gamma$, and $E_{CUT}$ with the rotational phase. The hydrogen column density ($n_{H}$) was fixed at the phase averaged value of $4.37 \times10^{22} \;cm^{-2}$. It is obvious that all the spectral parameters feature strong variations with a large amplitude over the pulse phase. The flux of the source varied between ~(3-5)$\;\times\;10^{-11}\;erg\;cm^{-2}\;s^{-1}$ and follows the continuum pulse profile as shown in the background in grey color (see Figure 8). The Photon Index ($\Gamma$) having maximum and minimum values of 1.29 in the phase interval (0.5-0.6) and 0.97 in the phase interval (0.3-0.4). The cutoff energy ($E_{c}$) parameter of the \textsc{cutoffpl} model was seen to highlight variability at all phases having maximum value of 29.83 keV in the phase interval (0.2-0.3) and minimum value of 12.03 keV in the initial phase. 

\section{Discussion and Conclusion}
In this paper, we have investigated the BeXRB IGR J21347+4737 in the energy band 0.5-50 keV. Previous work of (Reig \& Zezas 2014) have shown the timing and spectral properties of the source in the 1-12 keV energy range by analyzing the XMM-Newton observation. The timing analysis of the light curve detects coherent pulsation of the source at 322.738$\;\pm\;0.018$ s. In the same analysis using XMM-Newton observation, we found the spin period of the source at 320.351$\pm0.022$ s. This leads to the conclusion that the source has undergone spin down over the span of more than 7 years by 0.341 s $year^{-1}$. When the source pulsation was first discovered, it was in a low optical state meaning when the $H_{\alpha}$ line was in absorption, which implies the disappearance of the Be star's circumstellar disc (Reig \& Zezas 2014). However, the detection of pulsations in both the XMM-Newton and the NuSTAR observation indicates that the accretion mechanism still remains active. The main contribution to X-ray emission in the absence of disc is the accretion from a stellar wind (Reig \& Zezas 2014).

The pulse profile of IGR J21347+4737 in different energy bands demonstrate a single-peak, asymmetric in nature, indicating that the pulsar is probably turned towards the observer by one of its poles and leaving the other one to be practically invisible. Various theoretical model formulations are there that explain the asymmetric nature of the pulse profiles. One of the formulations being distorted magnetic dipole field. In this formalism, the magnetic poles not being opposite to one another may be one of the probable reasons for the asymmetric shape of pulse profiles (Parmar et. al. 1989; Leahy
1991; Riffert et. al. 1993; Bulik et. al. 1995). The other reason being considered is due to an asymmetric accretion stream (Basko
\& Sunyaev 1976; Wang \& Welter 1981; Miller 1996). As evident from Figures 2 and 3, the profile is single-peaked reflecting a pencil beam pattern in the neutron star radiation. The simple single-peaked profile of X-ray pulsars in the broadband energy range 3-79 keV is supported by the argument of Vasilopoulos et al. (2017). The source luminosity in our spectral analysis is of the order of $10^{35}$ for the NuSTAR. At this luminosity, we cannot expect the formation of an extended accretion column (Mushtukov et al. 2015a,b). Therefore most of the X-ray photons in this scenario should be originating from the region very close to the surface of the neutron star. As a result, the emission pattern in
such a case is characterized by a pencil beam shaped (Basko \& Sunyaev 1976). Thus, it is convincing to say that the source may be accreting in the sub-critical regime. Generally, a pencil beam pattern is characterized by luminosities marked lower than the critical luminosity ($L_{c}$). $L_{c}$ is defined as the luminosity below which the accretion phenomenon is characterized by pencil beamed pattern and above which the accretion phenomenon is characterized by a fan-beamed pattern. The pencil beamed pattern indicates the fact that the source may be accreting in the sub-critical regime while the fan-beamed pattern indicates that the source may be accreting in the super-critical regime. In the sub-critical regime, the accreted matter reaches the surface of the Neutron star through nuclear collisions with atmospheric protons or the coulomb collisions with thermal electrons (Harding et al.
1994), whereby, the emission phenomenon occurs from the top of the column (Burnard Arons \& Klein 1991).
It is also clear that the observed source luminosity of NuSTAR is much greater than the minimum luminosity ($2.2\;\times\;10^{32}\;erg\;s^{-1}$ (Reig \& Zezas 2014) at which  the propeller effect sets in.

%A strong anti-correlation was noted between the flux and the HR over the pulse suggesting the dominant nature %of the fan beam pattern in the NS radiation. The fan beam pattern of the source is further supported by the %positive correlation between the POWERLAW photon index and  the energy flux over the pulse phase. 
The pulse profile of the source is represented in Figures 2 and 3 and evolves with time indicating a change in the accretion geometry. 

The PF of the source shows a dramatic variation with energy that is rarely observed in low-luminosity X-ray pulsars. Initially, the PF is found to decrease steadily with energy upto 15 keV followed by a non-monotic increasing trend above 15 keV which is typical for X-ray pulsars (Lutovinov \& Tsygankov
2009). Such variations are usually seen at higher luminosities but rarely observed at low luminosities.

 The broadband 0.5-50 keV spectra of the source are well fitted by an absorbed \textsc{cutoff power-law} modified with cutoff at high energy. Other continuum models in the form of \textsc{highecut} were tested in place of \textsc{cutoffpl} and the given combinations also fit the spectra very well. No other emission components in the form of blackbody or Fluorescent iron line were required for modifying the spectra. The detection of the iron line provides strong evidence for the presence of material in the vicinity of the source. In addition, no absorption lines in the form of the Cyclotron line were present in the phase-averaged spectra.  There was no significant presence of features like CRSF that leads to the straightforward way to determine the magnetic field in such systems. The lack of presence of the CRSF suggests that the magnetic field of the source is either weaker than $\sim5\times10^{11}\;G$ or stronger than $\sim6\times10^{12}\;G$ considering the lower and upper limits of the observatory NuSTAR full energy range (3-79) keV where the sensitivity allows us to non-detection of such features below and above the range. 
 %The estimated magnetic field is consistent with the works of Gorban et. al. 2022.

When a source undergoes an outburst, it quickly reaches its peak luminosity and then the rate of mass accretion slows down. After an outburst, the pulsar could go through a phase called the propeller phase, which causes the luminosity to suddenly decrease. $L_{prop}$ stands for the corresponding luminosity at which the source transitions to the propeller phase. One of the causes of the luminosity's abrupt decline is the magnetosphere's centrifugal barrier, which is produced by its faster-than-Kelvin-velocity rotation. The expression for the propeller luminosity ($L_{prop}$) is,

\begin{equation}
L_{prop}\;=\;4\times10^{37}k^{7/2}B_{12}^{2}P_{s}^{-7/3}M_{1.4}^{-2/3}R_{6}^{5}[erg/s]
\end{equation} 

where k=0.5 is used to account for the interaction between the magnetosphere and the accretion flow. (Ghosh \& Lamb 1978), $B_{12}$ is the magnetic field in units of $10^{12}G$, $M_{1.4}$ is the mass of the NS in units of solar mass and $R_{6}$ is the radius of the NS in units of $10^{6}$.
If the magnetic field of the source is constrained at $10^{14}\;G$, using  the source pulsation of $\sim$322 s, we estimated $L_{prop}\sim5\times10^{32}\;erg \;s^{-1}$. And considering the magnetic field of the source at $10^{12}$ G, we found $L_{prop}\sim5\times10^{30}\;erg \;s^{-1}$.Thus the calculated $L_{prop}$ is roughly about $10^{3}-10^{5}$ times less than the observed luminosity of the source. It is therefore reasonable to conclude that the source with a pulse period of 322 s and a magnetic field of $10^{12}-10^{13}\;G$ may not reach the propeller phase. As a result, the source's magnetic field must be in the order of $10^{14}\;G$ for it to enter the propeller phase at the measured luminosity of the source.

 \section*{Data availability}
 
 The observational data used in this study can be accessed from the HEASARC data archive and is publicly  available for carrying out research work.

\section{Acknowledgement}
This research work have utilized the NuSTAR data archived by the NASA High Energy
Astrophysics Science Archive Research Center (HEASARC) online
service which is maintained by the Goddard Space Flight Center. This work
has made use of the NuSTAR Data Analysis Software (NuSTARDAS)
jointly developed by the ASI Space Science Data Center (SSDC,
Italy) and the California Institute of Technology (Caltech, USA). We further acknowledge the use of
public data from XMM-Newton observatory. We would like to thank the anonymous reviewer for his/her kind suggestion which helped us in improving the manuscript in the present form.
%%use \tablenotes{footnote} to get the table foot note

%%Use table* environment to get the table spanning both the columns

%\begin{table*}[htb]
%\tabularfont
%\caption{Caption text here}\label{secondTable}
%\begin{tabular}{lccccccccccccr}
%\topline
%\textbf{head1}&\multicolumn{11}{c}{\textbf{head2}}&\textbf{head3}\\
%\midline
%one& two &three&four&five&six&seven&eight&nine&ten&eleven&twelve&thirteen\\
%1&2&3&4&5&6&7&8&9&10&11&12&13\\
%aaa&bbbb&cccc&ddddd&eee&ffff&ggggg&hhhhhhhh&iiii&kkkkkk&hhh&jjjjjj&lllll\\
%\hline
%\end{tabular}
%\tablenotes{Table footnote here. Table spanning both the columns.}
%\end{table*}

%%An example of a figure

%\begin{figure}[!t]
%\includegraphics[width=.8\columnwidth]{fig1.eps}
%\caption{caption goes here}\label{figOne}
%\end{figure}

%%An example of a double column figure
%%Use figure* environment

%\begin{figure*}
%\centering\includegraphics[height=.15\textheight]{fig1.eps}
%\caption{caption spanning two columns}
%\centering\includegraphics[height=.25\textheight]{fig1.eps}
%\caption{caption here}
%\end{figure*}

%%use \balance somewhere in the left column of the last page to balance the two columns in the end page

%%References section

\end{document}